\documentclass{raa}
\usepackage{lineno}   % 加行号包
\usepackage{graphicx,times}             %for PS/EPS graphics inclusion, new
\usepackage{natbib}
\usepackage{amssymb,amsmath}
\usepackage{comment}
\bibpunct{(}{)}{;}{a}{}{,}

\usepackage[pagebackref=true]{hyperref}

\begin{document}

  \title{Onboard catalog of known X-ray sources for EP-WXT}
%   \subtitle{I. Place Your Subtitle Here}

   \volnopage{Vol.0 (20xx) No.0, 000--000}      %%preserved for Editor. DOn't remove!
   \setcounter{page}{1}          %%starting page, preserved for Editor. DOn't remove!

   \author{He-Yang Liu %(刘禾阳) %% Put your Chinese name in "( )" if you like. Note to open line 11 "\usepackage[UTF8]{ctex}"
      \inst{1}
    \and Haiwu Pan
      \inst{1}
    \and Yuan Liu
      \inst{1}
    \and Hong-Bo Cai
      \inst{1}       
      \and Yuyin Tan
       \inst{2}
      \and Haitao Xu
       \inst{2}
       \and Changbin Xue
       \inst{2}
       \and Xiaojin Sun
       \inst{3}
       \and Yifan Chen
       \inst{3}
       \and Zhiwei Cheng
       \inst{3}
       \and Ying Li
       \inst{3}
       \and Yulong Xue
       \inst{3}
       \and Huaqing Cheng
      \inst{1}
      \and Xuan Mao
      \inst{1, 4}
      \and Zhixing Ling
      \inst{1}
      \and Chen Zhang
      \inst{1}
       \and Weimin Yuan
      \inst{1}      
   }

   \institute{National Astronomical Observatories, Chinese Academy of Sciences,
             Beijing 100101, China; {\it panhaiwu@nao.cas.cn, liuyuan@nao.cas.cn}\\
        \and
             National Space Science Center, Chinese Academy of Sciences, Beijing 100190, China\\
        \and
             The Shanghai Institute of Technical Physics of the Chinese Academy of Sciences, Shanghai 200083, China\\
        \and
             School of Astronomy and Space Science, University of Chinese Academy of Sciences, Chinese Academy of Sciences, Beijing 100049, China\\
\vs\no
   {\small Received 20xx month day; accepted 20xx month day}}

\abstract{
The Einstein Probe (EP) is dedicated to explore the dynamic X-ray universe and capture transient events in real time with its Wide-field X-ray Telescope (WXT). However, WXT's unprecedentedly large instantaneous field of view—exceeding 3,600 square degrees—simultaneously encompasses numerous known X-ray emitters. Distinguishing genuine novel transients from these persistent sources is a critical observational challenge. To resolve this, EP-WXT incorporates a dedicated reference catalog of known X-ray sources directly into its onboard data processing and triggering system. In this paper, we detail the compilation of this onboard catalog. By merging data from the ROSAT All-Sky Survey, the MAXI source list, and a curated stellar flare candidate catalog, we constructed a robust baseline database of 9,000 sources. This catalog provides coordinates, baseline count rates, and spatial veto radii for exceptionally bright emitters. Real-time cross-matching against this database effectively decouples known background sources from the transient alert stream. Recent in-orbit operations validate the high stability and efficiency of this catalog-driven trigger system. Notably, the catalog successfully masked over 6,100 potential triggers from known active stars. This proves its essential role in ensuring EP's rapid and accurate response to genuine astrophysical discoveries.
\keywords{instrumentation: miscellaneous --- telescopes --- catalogs --- X-rays: general}
}

   \authorrunning{H.-Y. Liu et al.}            %author_head in even pages
   \titlerunning{EP-WXT onboard catalog}  % title_head in odd pages

   \maketitle
%% The author head (on even pages) and the title head (on odd pages) will be
%% automatically extracted from \author{} and \title{}. Whenever the title is too long,
%% you will be asked to supply a shorter one by inserting either \authorrunning{} or
%% \titlerunning{} before \maketitle. Anyway, you can specify your own heads.
%%
%%
%% Note: In the following text body of your manuscript, please note several differences from
%%       other major journals:
%% (1) \subsection{Please Capitalize the First Letter of Each Notional Word in Subsection Title}
%% (2) Please Capitalize the First Letter of Each Notional Word in all tables' captions

%
%________________________________________________ sections below
%
\section{Introduction}
\label{sec:Intro}

The X-ray sky is highly dynamic, hosting a wide variety of transient and variable phenomena over timescales ranging from milliseconds to years. These include gamma-ray bursts (GRBs), tidal disruption events (TDEs), supernova shock breakouts (SN SBOs), stellar flares, and accretion-driven outbursts from compact objects such as neutron stars and black holes. These phenomena probe extreme physical conditions, including strong gravity, relativistic jets, and magnetic reconnection, making them key to understanding high-energy astrophysical processes. Given that their fluxes can vary by several orders of magnitude, systematic wide-field monitoring is essential for their discovery and characterization \citep{Brandt1995, Gehrels2004, Komossa2015}.

Over the past few decades, time-domain X-ray astronomy has been significantly advanced by several landmark space missions. Instruments such as the \textit{Swift} Burst Alert Telescope \citep[BAT;][]{Barthelmy2005} and the Monitor of All-sky X-ray Image \citep[MAXI;][]{Matsuoka2009} have enabled continuous sky monitoring, yielding the discovery of numerous transient events and variable sources. Concurrently, recent missions like eROSITA \citep{Predehl2021} have delivered unprecedentedly deep all-sky surveys in the soft X-ray band. Despite these achievements, a critical observational gap remains: existing instruments generally suffer from either limited sensitivity to short-timescale events or an insufficient combination of a large field of view (FoV) and high observing cadence, making it difficult to efficiently detect faint and fast transients.

The Einstein Probe \citep[EP;][]{Yuan2022, Yuan2025} is specifically designed to bridge this gap and explore the dynamic X-ray universe. By systematically detecting transients and orchestrating rapid follow-up observations, EP combines wide-field monitoring with high soft X-ray sensitivity to dramatically expand the discovery space for rare and faint phenomena. Its primary scientific objectives encompass the detection of GRBs, TDEs, SN SBOs, X-ray binaries, as well as electromagnetic counterparts to gravitational-wave and neutrino events. To effectively capture these unpredictable phenomena, EP employs a high-cadence strategy, repeatedly scanning a substantial fraction of the sky.

To achieve these scientific goals, the primary instrument onboard EP, the Wide-field X-ray Telescope (WXT), employs innovative lobster-eye micro-pore optics \citep{Angel1979, Zhang2022}. This technology provides an unprecedentedly large instantaneous FoV of over 3600~deg$^{2}$ within the 0.5--4.0~keV energy band, allowing WXT to monitor vast celestial areas during each orbit. To ensure the timely discovery of unpredictable transient events, EP utilizes a sophisticated on-board processing and triggering unit. This autonomous system is tasked with identifying new X-ray transients and rapidly distributing alerts within minutes via the Beidou satellite navigation system and/or the Very High Frequency (VHF) network.

However, the exceptionally large FoV of WXT introduces a significant observational challenge. At any given moment, hundreds of known X-ray emitters---such as X-ray binaries, bright active galactic nuclei, supernova remnants, and active flare stars---are simultaneously present in the detector's view. Many of these sources exhibit intrinsic flux variability or stochastic outbursts that can closely mimic the signatures of genuine novel transients. Without a reliable reference for the baseline X-ray sky, the autonomous triggering algorithm would inevitably interpret significant flux variations from these known sources as new transient events, thereby dramatically increasing the false-trigger rate.

Consequently, the implementation of an onboard catalog of known X-ray sources is an operational necessity for EP-WXT to address this problem. By embedding a reliable reference database directly into the WXT Data Processing Unit, the flight software can instantly cross-match incoming photon clusters and effectively decouple persistent emitters from the transient alert stream. This autonomous veto mechanism ensures that the mission's rapid response capabilities are exclusively dedicated to the detection and follow-up of genuine, novel astronomical discoveries.

In this paper, we present the methodology for constructing the onboard catalog of known X-ray sources for EP-WXT and detail its operational implementation. We describe the selection of historical input catalogs, the estimation of source count rate, and how this catalog used to identify sources in the onboard detection algorithms and its in-flight performance. This paper is organized as follows: Section~\ref{sec2} provides an overview of the EP observation strategy and transient triggering system. Section~\ref{sec3} describes the detailed compilation and construction of the reference catalog; and Section~\ref{sec4} presents the catalog's implementation and performance, covering its data structure, its integration into the real-time triggering logic, the in-orbit performance and the case for autonomous onboard updates.

%========================
\section{EP Observation strategy and onboard triggering}
\label{sec2}

\subsection{Observation Strategy}
\label{sec:obs_strategy}
EP operates in a 593 km circular orbit with a 96.5-minute period. Its payload consists of two main instruments: WXT and the Follow-up X-ray Telescope (FXT; \citealp{Chen2020FXT}). WXT uses innovative lobster-eye optics to provide an exceptionally large FoV of over 3600 square degrees, allowing it to continuously monitor the dynamic X-ray sky. In contrast, FXT is a highly sensitive instrument designed to take a deeper look, precisely localizing and characterizing the transients discovered by WXT.

During routine operations, EP relies primarily on the WXT survey mode. In a single orbit, the WXT points in three distinct areas of the night sky with minimal overlap. This efficient strategy covers most of the night sky in just three orbits, ensuring that any given coordinate is visited several times a day. Through this systematic approach, the WXT survey mode achieves complete all-sky coverage within a six-month period.

The core strength of EP is its rapid response capability, driven by a real-time onboard data processing system. When WXT detects a new transient above a certain brightness threshold, the spacecraft autonomously triggers a follow-up observation. It quickly slews to point FXT at the target within just 3 to 5 minutes. Simultaneously, a preliminary alert---containing the coordinates, flux, and spectral hardness---is downlinked to the ground within minutes via the Beidou navigation system and/or the VHF network. These alerts are immediately shared with the global astronomical community to coordinate multi-wavelength follow-up campaigns.

Beyond these autonomous triggers, EP employs additional strategies to maximize its scientific output. Fainter transients that miss the onboard trigger threshold are identified later during ground data processing. Although this introduces a telemetry delay of a few hours, these events can still trigger internal Target of Opportunity (ToO) observations. Furthermore, EP actively responds to external alerts, such as gravitational-wave and neutrino events. If an external event has a poorly constrained position, WXT can first scan the error region to find the X-ray counterpart, providing refined coordinates to guide FXT. While standard ToO commands take about 3~hours to upload, urgent commands can be sent via the Beidou system in a few minutes, making EP a highly flexible and rapid-response node in time-domain astronomy.

\subsection{Onboard Data Processing and Triggering System}
\label{sec:odpts}
 The most important requirement for the primary observation strategy of EP is the real-time detection of transients on the fly. To achieve this, the spacecraft is equipped with a dedicated Onboard Data Processing and Triggering System (ODPTS). 
 Rather than waiting for an observation to finish, the onboard computer searches for transient events in real time, processing data across all 48 CMOS detectors over a range of timescales while the WXT observation is still ongoing.

The ODPTS employs a highly efficient algorithm to locate new sources. First, it continuously accumulates the detected X-ray photons into two one-dimensional (1D) positional histograms, projected along the horizontal and vertical directions. By identifying potential source candidates in these 1D histograms and matching their coordinates, the system quickly calculates possible 2D positions. The software then extracts and screens the photons around these specific 2D coordinates to confirm and evaluate the detected sources.

Once a list of sources is generated, the ODPTS compares them with the onboard source catalog to filter out known persistent emitters. If a new transient is confirmed, the system autonomously triggers an immediate FXT follow-up observation. Concurrently, a real-time alert message is downlinked to the ground. 
Upon receipt, these transient alerts undergo rapid manual screening in real time before being made public, allowing the worldwide astronomical community to initiate multi-wavelength follow-up campaigns.

The flight software and hardware of the ODPTS are exceptionally robust and fast. It has been demonstrated that the system can process the data from all 48 CMOS detectors in less than 10 seconds. For a typical 1,000-second exposure during a survey pointing, the system achieves a source detection sensitivity of approximately 1 mCrab. Furthermore, both ground experimental tests and in-flight observations confirm that the source location accuracy is around 3 arcminutes, providing highly reliable coordinates for subsequent observations.

%========================

%========================
\section{Catalog Construction}
\label{sec3}
The primary objective of the onboard catalog is to provide a highly reliable database of known X-ray emitters that the EP ODPTS can rapidly query. Due to the stringent constraints on onboard storage, it is crucial to carefully select the most suitable sources to establish a baseline for the X-ray sky. In this section, we detail how this reference sample was compiled.

\subsection{Input source catalogs}

The construction of the EP-WXT onboard reference catalog is primarily driven by the need for all-sky completeness and the prioritized monitoring of known bright sources. The foundational data for this catalog were synthesized from two major historical X-ray missions (see Table~\ref{tab:missions}). We utilize the \textit{ROSAT} All-Sky Survey (RASS; \citealt{Voges1999}) as the foundational baseline, as its limiting sensitivity of $\sim 10^{-13}~\text{erg cm}^{-2}\text{ s}^{-1}$ far exceeds the typical $10^{-11}~\text{erg cm}^{-2}\text{ s}^{-1}$ threshold of a standard WXT 1200~s exposure. This ensures the catalog remains sufficiently deep and robust for the onboard trigger algorithm. This foundational database is further augmented by the MAXI source list \citep{Matsuoka2009}, which incorporates the most recent and brightest X-ray emitters \footnote{We have also verified the brightest sources in the 4XMM-DR12 catalog \citep{Webb2020} with a flux greater than $1 \times 10^{-10}$~erg~s$^{-1}$~cm$^{-2}$ in the $0.5\text{--}4.5$~keV band and multiple detections; no such source has been missed in the onboard reference catalog.}. By cross-matching these major historical catalogs and removing duplicates, the onboard system can assign higher matching priorities to prominent active sources, ensuring an efficient real-time identification process.

To optimize the allocation of limited mission resources, the catalog incorporates a dedicated subset of known stellar flare candidates, such as RS~CVn-type binaries, T~Tauri stars, and active late-type stars. Due to WXT’s vast FoV and high sensitivity, these magnetically active stars are frequently monitored and tend to trigger a large number of onboard alerts and subsequent autonomous follow-up observations. Without proper identification, this high frequency of stellar events could overwhelm the system's processing capacity and potentially interfere with the timely triggering and follow-up of other high-priority transients, such as the prompt emission of GRBs or SN SBOs. By integrating these known active stars into the reference database, the onboard software can effectively filter out routine stellar activity, ensuring that the mission's primary scientific targets receive priority.

A critical requirement of this catalog is to enable the ODPTS to estimate the expected count rate of each source across the EP-WXT energy band. Because we adopted RASS as our foundational database, the baseline count rates for all sources are initially provided in the RASS band (0.1--2.4 keV). To resolve this bandpass difference, the onboard software dynamically converts these RASS count rates into expected WXT count rates using an empirical relation, 
$C_{\text{ROSAT}} = \frac{750}{13} C_{\text{WXT}}$, where $C_{\text{ROSAT}}$ represents the count rate in the ROSAT PSPC band ($0.1\text{--}2.4\text{~keV}$),  $C_{\text{WXT}}$ is the corresponding expected count rate in the EP-WXT band ($0.5\text{--}4.0\text{~keV}$), and the scaling factor of $750/13$ is determined based on the ratio of the Crab Nebula count rates estimated for ROSAT \citep{Kirsch2005} and EP-WXT, with the latter derived from simulations. We note that employing a single scaling factor introduces systematic uncertainties due to variations in the intrinsic X-ray spectral shapes of different sources. Generally, the conversion factor depends on the spectral hardness: a softer spectrum corresponds to a lower WXT count rate for a given RASS count rate, whereas a harder spectrum corresponds to a higher one.  To quantitatively evaluate this effect, we performed a series of simulations using an absorbed power-law model\footnote{The systematic trend remains independent of the choice of specific spectral models (such as Blackbody, APEC, or Power-law), here we choose power law as adopted in the RASS spectral analysis of Crab \citep{Kirsch2005}.}, grid-searching a parameter space with $N_{\text{H}}$ ranging from $0$ to $1 \times 10^{22}\text{~cm}^{-2}$ and $\Gamma$ from $1.2$ to $2.5$. The simulation results indicate that the systematic discrepancy is generally constrained within $\sim 0.5\text{~dex}$, except for extreme cases of severe absorption or exceptionally soft spectra. Therefore, the estimated WXT count rates remain well within the acceptable tolerance and exert negligible impact on the efficiency of the onboard transient triggering algorithm.

%               one-column-spanning table
%________________________________________ Table 1: Use_of_the routines
\begin{table}
\begin{center}
\caption[]{X-ray satellites adopted as the main source catalogs in this work}
\label{tab:missions}

%%Please Capitalize the First Letter of Each Notional Word in table's caption

 \begin{tabular}{ccccc}
  \hline\noalign{\smallskip}
  Mission & Period of operation & Energy range & spatial resolution &  sensitivity @ 20~min  \\
    &  & (keV) &  & (erg s$^{-1}$ cm$^{-2}$) \\
  \hline\noalign{\smallskip}
  ROSAT & 1990-1999 & 0.1-2.4 & $\sim 30''$ & $\sim 1\times$ 10$^{-13}$ \\
MAXI/GSC & 2009–... & 2–20 & 1.5$^\circ$ & $\sim 1\times$ 10$^{-10}$\\
  \noalign{\smallskip}\hline
\end{tabular}
\end{center}
\end{table}

\subsubsection{Sources from RASS catalogs}

The \textit{Röntgen Satellite} (ROSAT), is a joint mission between Germany, the United States, and the United Kingdom, launched in 1990 and operated until 1999 \citep{Trumper1983}. Equipped with a Wolter-I grazing incidence telescope, ROSAT performed the first imaging X-ray all-sky survey in the soft X-ray band (0.1--2.4~keV). This mission revolutionized our understanding of the high-energy universe by identifying over 100,000 sources, ranging from stellar corona to distant active galactic nuclei (AGN). The RASS data has been published in two primary catalogs: the First ROSAT All-Sky Survey \citep[1RXS;][]{Voges1999} and the Second ROSAT All-Sky Survey \citep[2RXS;][]{Boller2016}. The 2RXS catalog utilizes more advanced point-source detection algorithms and improved background modeling, generally yielding a more accurate and complete sample of X-ray sources compared to its predecessor. However, this increased accuracy comes at the cost of spatial coverage. To maintain the reliability of the point-source detection process, the 2RXS pipeline intentionally masked several sky regions dominated by complex and strong extended emission. Notably, regions encompassing prominent supernova remnants such as the Vela SNR, Cassiopeia A (Cas A), and the Cygnus Loop were excluded. As a result, the 2RXS catalog suffers from incomplete all-sky coverage and critically misses information on several exceptionally bright soft X-ray emitters, which are quintessential sources capable of triggering the WXT onboard system.

To guaranty the maximum comprehensiveness required for the EP-WXT onboard catalog, we adopted a cross-matching strategy that integrates the information from both catalogs. By combining the high-precision measurements of 2RXS with the missing bright sources retrieved from 1RXS, we compiled a highly complete and reliable master sample of historical soft X-ray sources. 

\subsubsection{Sources from MAXI/GSC catalog}

The Monitor of All-sky X-ray Image (MAXI; \citealt{Matsuoka2009}) is an X-ray mission mounted on the Exposed Facility of the Japanese Experiment Module aboard the International Space Station. Launched in July 2009, MAXI is equipped with two sets of X-ray slit cameras: the Gas Slit Camera (GSC), which operates in the $2-30$~keV energy range, and the Solid-state Slit Camera (SSC), covering the $0.5-12$~keV band. Since its commencement of operations, MAXI has been continuously scanning the entire sky at a high cadence. Its sustained observational campaign has been instrumental in detecting novel X-ray transients and monitoring the long-term flux variability of known X-ray emitters. Over a decade of operation, MAXI has compiled a robust observational record of hundreds of bright X-ray sources. Given their persistent brightness and variability, these sources are highly relevant to the EP-WXT operational bandpass and have consequently been incorporated into the baseline onboard catalog. 

To compile the MAXI/GSC sub-catalog, we obtained the official reference source list from the MAXI website\footnote{\url{http://maxi.riken.jp/top/slist.html}}. The list was filtered by astronomical classification to select Active Galactic Nuclei (AGNs), X-ray binaries, and active stars. We evaluated the historical MAXI count rates of these candidates to confirm their general detectability by WXT, yielding a refined sample of 337 sources. These count rates were converted into equivalent ROSAT values via an empirical scaling factor anchored to the Crab Nebula, ensuring compatibility with the RASS-based foundational catalog.

\subsubsection{Compilation of the Stellar Flare Candidate Catalog}

Beyond steady X-ray emitters, a significant fraction of the variable X-ray sky is dominated by stellar flares, primarily originating from RS~CVn-type binaries, T~Tauri stars, and active M and K dwarfs in the local universe. Based on our simulation results, stellar flares are expected to be the most numerous transient sources detected by the EP mission, dominating the population of triggered events. These sources frequently exhibit coronal flares characterized by rapid rises and exponential decays, which can trigger the onboard alert system and initiate autonomous follow-up sequences. Given their expected dominance in the detected transient sample, it is highly necessary to incorporate a dedicated subset of known stellar flare candidates into the onboard catalog. 

The stellar flare candidate catalog is compiled from three independent sub-catalogs, prioritized based on historical occurrence rates observed by previous wide-field missions. During its first seven years of operation, the MAXI mission detected 106 flares across 27 flaring low-mass stars, a sample dominated by 14 RS Canum Venaticorum (RS CVn) systems and 9 dMe stars \citep{Tsuboi2017}. This indicates that these two stellar classes are the most frequent producers of prominent, high-luminosity X-ray flares in the solar neighborhood. Given that RS CVn systems represent the most active subclass within the broader family of Chromospherically Active Binaries (CAB), and that flaring dMe stars are practically synonymous with UV Ceti-type variables, we assign the highest priority to the CAB and UV Ceti catalogs to maximize the filtering of these prevalent transients. A supplementary catalog derived from the RASS combined with Gaia data serves as the third priority tier. The detailed compilation procedures are described below.

\textbf{Chromospherically Active Binaries (CAB).}
The CAB sub-catalog is derived primarily from the works of \citet{Eker2008} and \citet{Dempsey1993}. \citet{Eker2008} expanded the original CAB database to 409 sources, encompassing both RS~CVn and BY Draconis (BY~Dra) variables. In parallel, \citet{Dempsey1993} compiled a sample of 136 RS~CVn systems detected during the RASS, 135 of which have confirmed positional information in the SIMBAD database. Cross-matching these two samples reveals significant overlap, yielding 9 unique additional sources from the \citet{Dempsey1993} list. The merged catalog therefore consists of 418 unique CAB sources.

To estimate the expected baseline X-ray photon count rate for each source in the EP-WXT bandpass, we cross-matched the 418 CAB sources with the 2RXS \citep{Boller2016} and the Second ROSAT Pointed Observations Catalog \citep[2RXP;][]{ROSAT2000} using a search radius of $15''$. This yielded 279 matches in 2RXS and 77 matches in 2RXP. For sources detected in both catalogs, the count rate derived from the deeper and more precise 2RXP was prioritized; otherwise, the 2RXS count rate was adopted. 

For sources absent from both ROSAT catalogs, the expected count rates were estimated indirectly. First, utilizing the matched 2RXS population, we derived an empirical conversion factor between the RASS count rate and the unabsorbed X-ray flux (modeled assuming a MEKAL thermal plasma). Subsequently, we calculated the quiescent X-ray flux for the undetected sources using the X-ray luminosities provided by \citet{Eker2008} and the corresponding distances queried from the SIMBAD database. For sources lacking a reported luminosity, a conservative baseline luminosity of $2 \times 10^{30}~\mathrm{erg~s^{-1}}$ was assumed. Finally, the derived fluxes were converted to the expected count rates using the aforementioned empirical conversion factor.

\textbf{UV Ceti-type Flare Stars.}
To construct the UV Ceti sub-catalog, we merged two specialized samples: the comprehensive flare star database by \citet{Gershberg1999} and the binary-focused catalog by \citet{Tamazian2014}. The \citet{Gershberg1999} catalog serves as a foundational repository for 463 UV Ceti-type and related flare stars in the solar vicinity, heavily incorporating historical optical flare monitoring as well as quiescent soft X-ray luminosities derived from the RASS (2RXS). To complement this, we integrated 138 sources from \citet{Tamazian2014}, which specifically catalog UV Ceti-type flare stars residing in nearby visual binary systems (typically within 25 parsecs) and provides updated astrometric and astrophysical parameters. 

These two samples were cross-matched using a search radius of $5''$ to identify and remove duplicates. This automated step was supplemented by careful manual verification for coordinate pairs with slightly larger angular separations, which frequently arise from the high proper motions of these nearby dwarfs or their binary orbital displacements. This merging process yielded a final sub-catalog comprising 516 distinct UV Ceti flare stars. The expected baseline X-ray count rates for these 516 sources were then determined utilizing the same methodology as described for the CAB catalog.

\textbf{ROSAT-Gaia Flare Candidates.}
To enhance the completeness of the stellar flare catalog, we developed a supplementary list of potential flare candidates by integrating the 2RXS catalog \citep{Boller2016} with stellar parameters from the \textit{Gaia} Data Release 2 (DR2; \citealp{Gaia2018}). Based on our sensitivity simulations, stellar flares originating from distances exceeding 1~kpc are generally below the detection limit of EP-WXT with an exposure of 1200~s. Therefore, we restricted the initial \textit{Gaia} DR2 sample to stars within 1~kpc, resulting in approximately 19.58 millions of targets. 

Following the empirical scaling relations provided by \citet{Guarcello2019}, we extrapolated the optical luminosity during a white-light flare from the star's quiescent optical luminosity, which subsequently allowed for the estimation of the peak X-ray flare luminosity. By combining this estimated luminosity with the \textit{Gaia}-derived distance, we calculated the anticipated peak X-ray flux for each candidate. Sources with an expected peak flux below $10^{-12}~\mathrm{erg~cm^{-2}~s^{-1}}$ were discarded, leaving 154,356 energetic candidates. Finally, to verify the presence of intrinsic X-ray activity, we cross-matched these candidates with the 2RXS catalog using a $15''$ search radius, yielding a final filtered sample of 4,717 supplementary sources.

It is important to note two caveats regarding this supplementary catalog. First, unlike the CAB and UV Ceti catalogs, which utilize precise optical/infrared coordinates from the literature, the coordinates for this sub-catalog adopt the X-ray positions listed in the 2RXS catalog. Second, owing to the relatively large positional uncertainties of the RASS, a fraction of these 4,717 cross-matches may be spurious (e.g., a background active galactic nucleus coincidentally aligned with a foreground \textit{Gaia} star). Nevertheless, their inclusion serves as a conservative measure to reduce frequent, low-priority triggers, thereby preventing these events from interfering with the core scientific objectives of the EP mission.

\subsubsection{Extremely Bright/Extended Sources}
\label{sec:bright_sources}

Due to the unique cruciform point spread function (PSF) of the lobster-eye optics, extremely bright X-ray sources can produce prominent ``cross arms'' that extend across the CMOS detectors. These features, along with diffuse emission from large extended sources, can be misinterpreted by the ODPTS as multiple genuine new transients, leading to a high rate of false triggers.  

\begin{figure}
\centering
\includegraphics[width=\textwidth, angle=0]{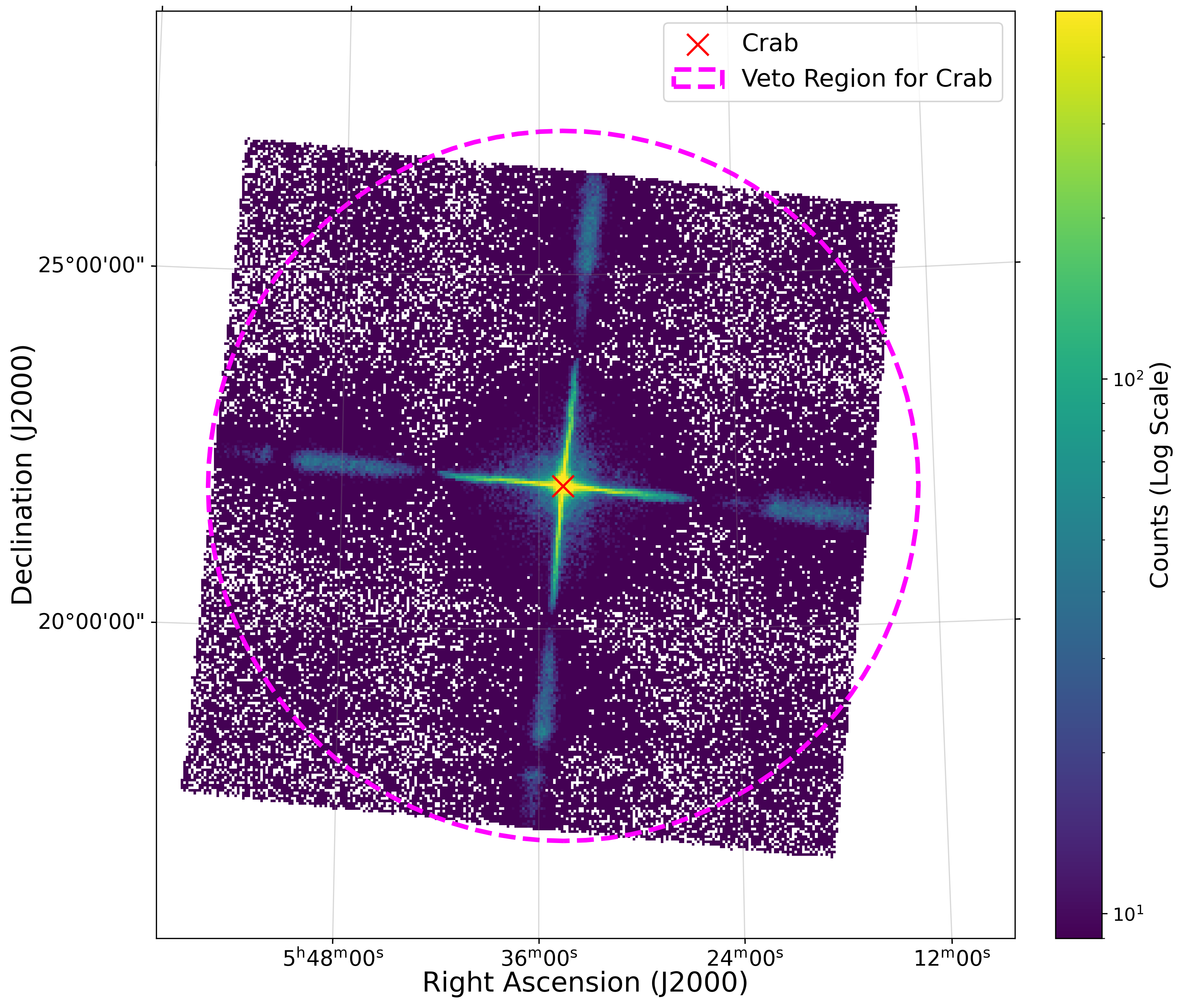}
\caption{This figure shows the point spread function (PSF) shape of the Crab Nebula as observed by EP-WXT. Due to the unique "lobster-eye" optics, such bright sources produce a prominent cruciform pattern that can trigger false alarms. To prevent this, a predefined veto region (the magenta dashed circle, with a radius of 5$^\circ$) is established around the source, within which any newly detected transients are automatically ignored by the onboard system.}
\label{Fig0}
\end{figure}

To mitigate this effect, these objects are specially handled and flagged within the onboard catalog. Each of these sources is assigned a specific angular distance, which serves as a veto radius. In the catalog, the ``Flag'' column records this distance in units of degrees. If the ODPTS detects a potential transient that falls within this specified radius from a known bright or extended source, the onboard triggering system will automatically ignore the event. An example of this handling is illustrated in Figure~\ref{Fig0}, which displays the prominent cruciform PSF of the Crab Nebula and its assigned veto region in the radius of $5^\circ$. Currently, 10 such sources have been identified and included in this special category, the details of which are provided in Table~\ref{tab:bright_source}.

The specific veto radius for each source is determined based on its spatial morphology observed with EP-WXT. For bright point sources (such as Sco~X-1 and Crab), the radii are derived from pre-launch simulations. In general, a higher X-ray flux produces more prominent and extended cruciform PSF arms, thereby requiring a larger veto radius to fully enclose the scattering artifacts. For extended sources (e.g., the Cygnus~Loop), the radii are adopted from their physical angular sizes in the X-ray band to ensure that the diffuse emission is completely masked. In addition, a radius of $1^\circ$ represents the absolute minimum constraint accepted by the onboard triggering algorithm for these flagged objects. Therefore, even for smaller or more compact bright sources like IC~443, the veto radius is capped at a minimum of $1^\circ$. Note that this $1^\circ$ minimum constraint applies exclusively to the bright or extended sources designated in Table 2; for standard sources in the reference catalog, a default veto radius of $6^\prime$ is applied.

Furthermore, beyond the static veto region, the ODPTS algorithm incorporates a dynamic check to identify spurious transients caused by the cruciform arms of unflagged sources. Upon detecting a new transient candidate, the software automatically evaluates the photon count rates along its expected cross-arm directions. These values are then compared directly with the count rate of the candidate itself. If the count rate along any of the arms is higher than that of the candidate's core, the algorithm deduces that the detection is merely a PSF artifact rather than a genuine astrophysical source and consequently discards it.

\subsection{Astrometric Considerations}
 To accommodate WXT’s typical localization accuracy of 3~arcminutes, coordinates from historical catalogs were optimized and formatted into a simplified grid to accelerate the onboard spatial cross-matching algorithm. The positional accuracy of RASS is approximately 30~arcseconds, which adequately fulfills the requirements for the onboard catalog. In contrast, since the positional precision of MAXI/GSC is relatively coarse (limited by its $\sim 1.5^\circ$ spatial resolution; \citealp{Sugizaki2011}), the source positions adopted in the onboard catalog for these objects are the best available coordinates reported in SIMBAD \citep{Wenger2000}. For the stellar flare sample, coordinates are obtained from the Gaia Data Release 2 \citep[DR2;][]{Gaia2018}. Overall, the coordinate precision provided in the onboard catalog is well-suited to the operational requirements of the ODPTS algorithm.

\subsection{Construction of the Final Catalog}
\label{sec:final_catalog}

The final onboard catalog is built through a step-by-step merging process. This approach ensures maximum reliability for the X-ray sky while strictly meeting the satellite's memory and data limits. The catalog is created by merging and filtering several specific source groups based on their priority. 

First, we give the highest priority to the most critical known sources. These include the 10 extremely bright and extended sources, bright sources in RASS catalog with count rate great than $ 1~\text{counts~s}^{-1}$, the 337 sources from the MAXI/GSC list, and all the compiled stellar flare candidates (CAB, UV~Ceti, and ROSAT-Gaia samples). These sources are included first because they are highly variable or bright enough to affect onboard triggering.

To fill the remaining capacity of the database up to the predefined limit, we utilize the extensive RASS catalog. The RASS entries are initially cleaned and sorted in descending order based on their recorded ROSAT count rates to prioritize the most significant and robust persistent X-ray emitters. 
During this ranking and pre-selection process, a conservative filtering threshold of $\ge 0.001~\text{counts~s}^{-1}$ is applied to eliminate exceptionally faint or low-significance background fluctuations that fall well below the typical real-time detection threshold of a standard WXT exposure.

After removing duplicate sources caused by overlaps between different catalogs, we then select the top-ranked sources one by one until reaching the memory limit. This process gives a final onboard catalog of 9,000 unique sources. It provides a optimal and robust reference baseline for the ODPTS to perform real-time transient searching, identification, and filtering. 

\begin{table}
\begin{center}
\caption[]{Bright sources of the EP onboard catalog}
\label{tab:bright_source}

 \begin{tabular}{lcccc}
  \hline\noalign{\smallskip}
    Name & RA & Dec & Count Rate & Flag \\
    & (deg) & (deg) & count s$^{-1}$ & (deg) \\
  \hline\noalign{\smallskip}
 Coma Cluster & 194.9350 & 27.9125 & 1.9 & 1 \\
 Crab & 83.6324 & 22.0174 & 750 & 5 \\
 Cyg X-2 & 326.1715 & 38.3214 & 300 & 5 \\
 Cygnus Loop & 312.7500 & 30.6667 & 1000 & 5 \\
 IC 443 & 94.2500 & 22.5700 & 0.62 & 1 \\
 Perseus Cluster & 49.9482 & 41.5150 & 5 & 1 \\
 RCW 103 & 244.3875 & $-$51.0333 & 4.95 & 1 \\
 RCW 86 & 220.1208 & $-$62.6450 & 1.67 & 1 \\
 Sco X-1 & 244.9795 & $-$15.6403 & 7500 & 15 \\
 Vela & 128.5000 & $-$45.8333 & 2000 & 5 \\
  \noalign{\smallskip}\hline

\end{tabular}

\begin{description}
\item[\textsuperscript{a}] Note: The \textit{Flag} column represents the angular distance (in units of degrees) used as a veto radius. If a newly detected transient is within this distance from the known bright source, it will be automatically ignored by the onboard triggering system.
\end{description}

\end{center}
\end{table}

\section{Catalog Implementation and Performance}
\label{sec4}

\subsection{Catalog Structure}
\label{sec:catalog_structure}

The onboard reference catalog has been rigorously streamlined to ensure highly efficient real-time spatial retrieval under limited onboard computational resources. The catalog is primarily composed of three physical quantities: celestial coordinates (Right Ascension, and Declination, Dec) and the baseline X-ray count rate in the ROSAT $0.1-2.4~\mathrm{keV}$ energy band. The RA and Dec are expressed in decimal degrees (J2000 epoch). The count rate, expressed in $\mathrm{count~s^{-1}}$, defines the expected X-ray emission level of the source and serves as the baseline flux for the onboard real-time triggering algorithm to evaluate sudden brightness enhancements. Notably, the sub-catalog for extremely bright X-ray sources incorporates an additional parameter, \textit{Flag}, which specifies the veto radius in degrees (see Table~\ref{tab:bright_source}). This parameter instructs the onboard software to automatically veto any transient candidates detected within the vicinity of bright, persistent X-ray sources (e.g., Sco X-1, Crab) to prevent frequent spurious triggers.

Figure~2 illustrates the spatial distribution of the 9,000 sources comprising the final onboard catalog. Within the equatorial coordinate system, the sources demonstrate a relatively isotropic distribution, thereby ensuring comprehensive all-sky coverage. Specifically, taking into account the veto radius assigned to each source (nominally $6'$ for standard sources and larger for specific bright ones), the cumulative masked region accounts for only $\sim 3$\% of the entire celestial sphere. Such a minimal footprint guarantees that the onboard triggering algorithm remains highly sensitive to novel transient events across the vast majority of the sky.
Furthermore, Figure~3 presents the statistical distribution of the expected ROSAT count rates for these sources. The count rates span several orders of magnitude, extending from $10^{-3}~\mathrm{count~s^{-1}}$ to in excess of $10^{4}~\mathrm{count~s^{-1}}$. A prominent peak is observed between $0.1$ and $1~\mathrm{count~s^{-1}}$, which seems to characteristic of the typical quiescent soft X-ray emission of active stellar coronae in the solar neighborhood. By structuring the catalog around these essential parameters, the onboard data processing unit can rapidly compute the angular separations between incoming transient candidates and known persistent sources. Simultaneously, it evaluates their observed fluxes against the established baseline count rates to accurately classify and filter transients.

\begin{figure}
\centering
\includegraphics[width=\textwidth, angle=0]{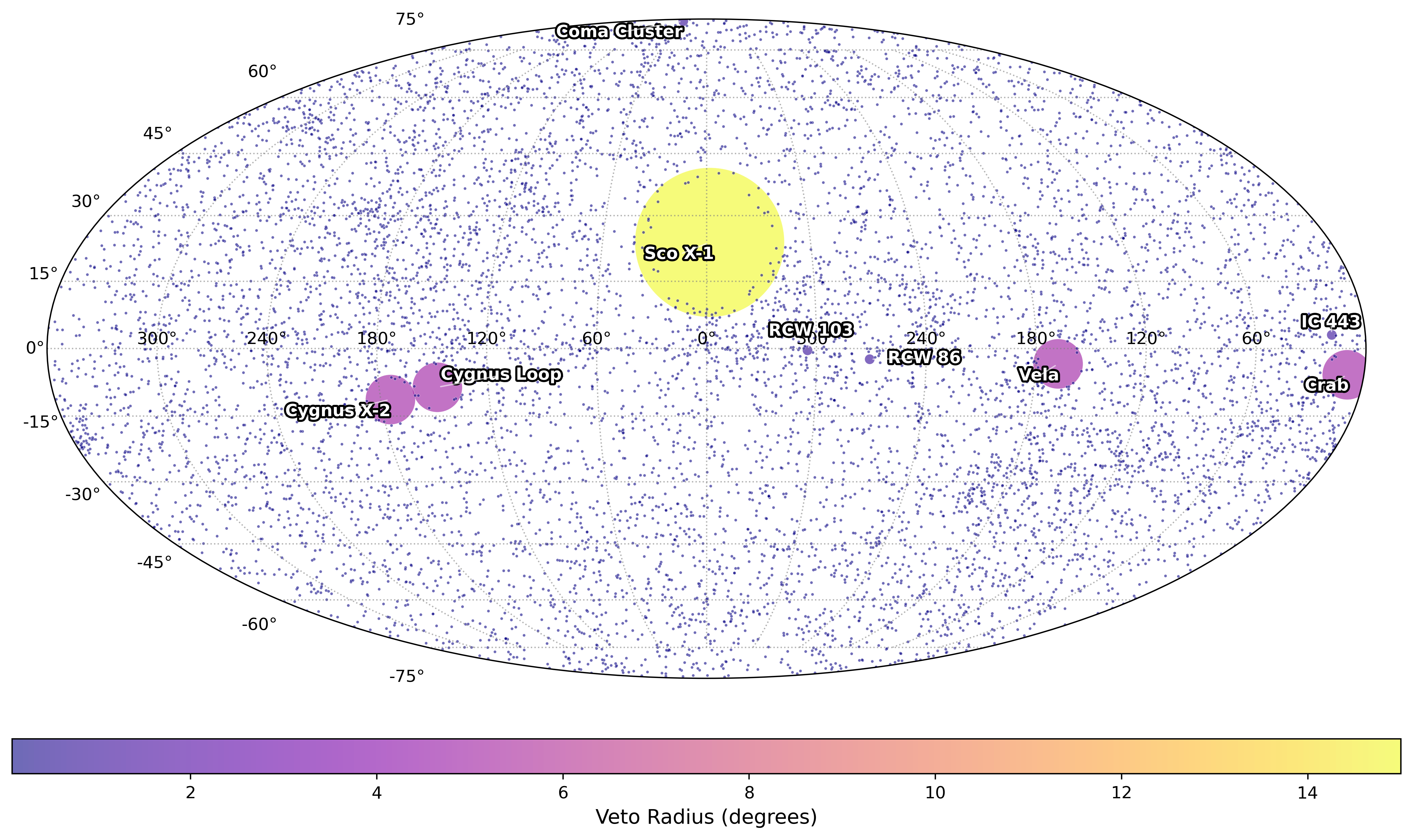}
\caption{Distribution of the sources of the EP-WXT onboard catalog on the sky in Galactic coordinates.}
\label{Fig1}
\end{figure}

% 考虑Nominal count rate?
\begin{figure}
\centering
\includegraphics[width=\textwidth, angle=0]{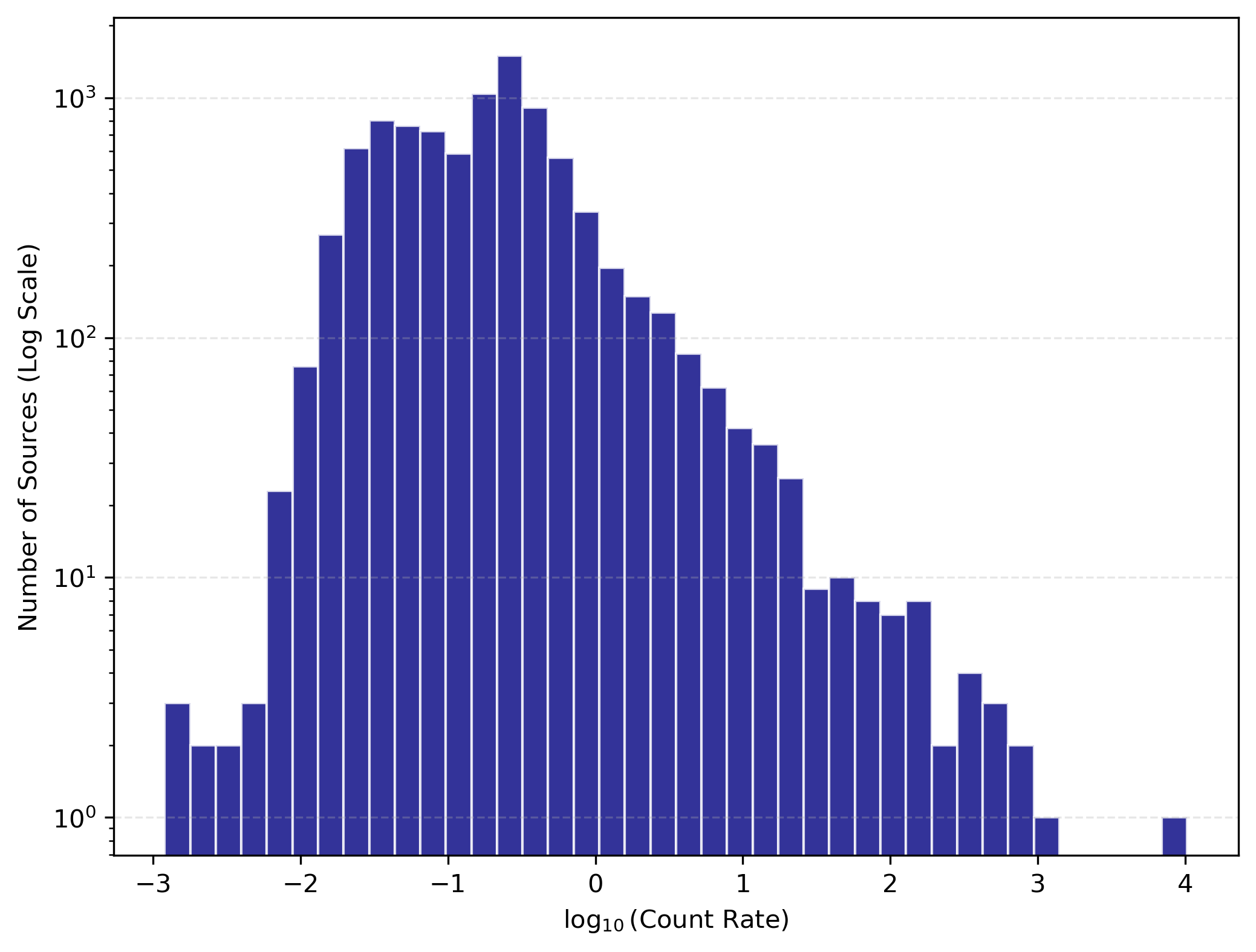}
\caption{Count rate distribution of the sources of the EP-WXT onboard catalog; the count rates has been transformed into ROSAT ones in the band of 0.1--2.4~keV.}
\label{Fig2}
\end{figure}

\subsection{Usage of the Onboard Catalog}
\label{sec:catalog_usage}

The static reference catalog is stored in the non-volatile memory of the WXT Data Processing Unit (DPU). The primary EP observation strategy demands the real-time detection of transients on the fly, a task executed by the dedicated ODPTS. During routine survey observations, the onboard software executes the following sequential trigger logic:

\textbf{1. Real-time Source Extraction:} 
As described in Sec~\ref{sec:odpts}, the ODPTS continuously processes data from all 48 CMOS detectors over various timescales while the observation is ongoing. It employs a highly efficient projection algorithm to rapidly extract and localize active X-ray sources, achieving a detection sensitivity of approximately $1~\mathrm{mCrab}$ for a typical 1,000-second exposure with a localization accuracy of roughly $3'$.

\textbf{2. Cross-Matching and Variability Check:} 
Once a list of active sources is generated, the ODPTS compares their coordinates against the onboard reference catalog to filter out known persistent emitters. If the position of a newly detected source falls within the defined spatial error radius of a cataloged source, it is flagged as ``known.'' For these cataloged sources, the software then compares the currently detected count rate with the expected reference count rate. An alert is only triggered if the observed flux exceeds the historical baseline by a predefined threshold of 10 times\footnote{This threshold is adjustable via ground telecommands to adapt to in-orbit performance.}, indicating a significant outburst or a powerful stellar flare.

\textbf{3. Uncataloged Transients and the Temporal Catalog:} 
If a detected source has no counterpart in the onboard reference catalog, it is flagged as a new transient candidate. In such cases, the system autonomously triggers an immediate Follow-up X-ray Telescope (FXT) observation, and a quick alert packet is queued for immediate downlink to the ground.

Crucially, to prevent the onboard software from repeatedly triggering on the same newly discovered transient during subsequent integration timescales or adjacent pointings, these uncataloged sources are dynamically registered into an onboard \textit{temporal catalog}. This temporal catalog functions as a temporary circular buffer operating on a First-In-First-Out (FIFO) basis. Once the allocated memory for the temporal catalog is fully saturated, the oldest transient entries are automatically overwritten by the newest detections, ensuring that the system remains highly responsive to pristine outbursts without being overwhelmed by previously reported events.

\subsection{In-flight Performance}
Since the launch of the EP on January 9, 2024, the onboard catalog has been fully integrated into the real-time data processing pipeline of the WXT. During the initial in-orbit operations, the catalog has demonstrated high stability and efficiency in cross-matching newly detected X-ray signals against known sources.

\begin{figure}
\centering
\includegraphics[width=\textwidth, angle=0]{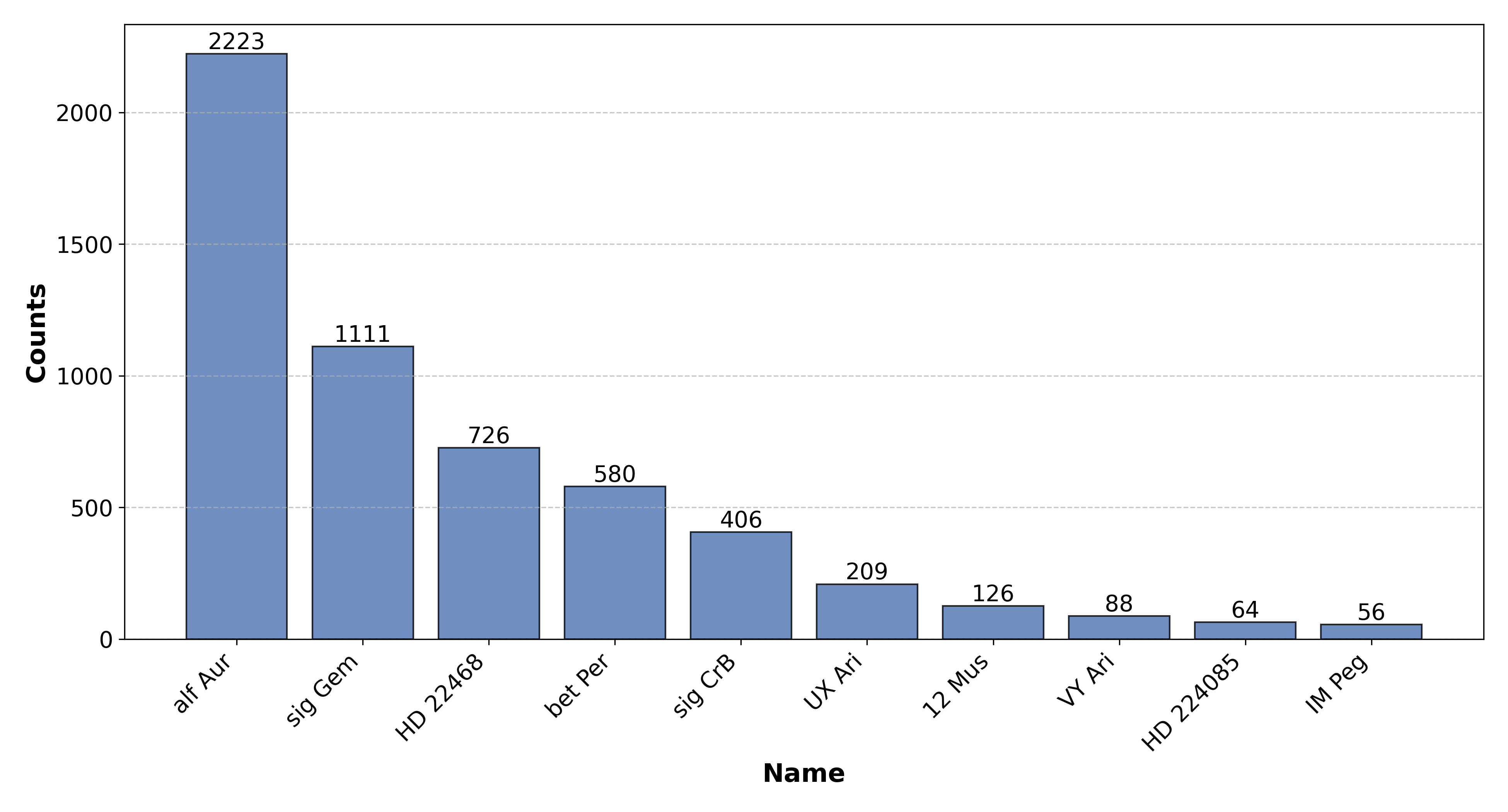}
\caption{Distribution of the top 10 potential stellar triggers successfully mitigated by the EP onboard catalog from January 9, 2024, to April 30, 2026. The vertical axis represents the total number of occurrences that met the flux criteria but were effectively masked by the catalog, thereby preventing these alerts.}
\label{Fig3}
\end{figure}

The primary objective of this catalog is to mitigate alerts caused by known sources, especially recurrent stellar flares. Actual operational results confirm that the spatial masking strategy—filtering out new detections that fall within the predefined exclusion radii of cataloged active stars—is highly successful in practice. According to ground-based data analysis spanning from January 9, 2024, to April 30, 2026, a total of 6,121 potential triggers originating from 85 cataloged stars were successfully masked (see Figure~\ref{Fig3} for the distribution of the top 10 potential stellar triggers). During the same period, only approximately 100 unmasked stellar flare triggers occurred onboard, maintaining an alert rate well within acceptable operational margins. 

This systematic filtration is vital for mission operations, as it prevents the limited real-time downlink bandwidth (e.g., via the BeiDou network) from being saturated by routine stellar activity. Furthermore, it crucially prevents the FXT from autonomously executing follow-up observations of these known flares. Consequently, the catalog ensures that the satellite's computational resources, instrumental pointing time, and rapid-alert capabilities are optimally preserved for genuinely serendipitous and unprecedented astrophysical phenomena.

\subsection{Onboard Updates}
\label{sec:onboard_updates}

The X-ray sky is intrinsically dynamic, and the population of active persistent sources evolves over time. To maintain optimal triggering efficiency and maximize scientific yield, the EP-WXT onboard catalog is designed to be fully modifiable rather than hard-coded into the flight software. As the mission progresses, the entire catalog—or specific entries—can be regularly updated, refined, or purged via telecommand uplinks from the ground station.

This update capability allows the reference catalog to be dynamically refined based on real in-flight operations and scientific requirements. For instance, during recent operations, we removed a group of historical X-ray binaries (XRBs) and nova from the catalog. This ensures that the WXT can successfully trigger observations if these volatile sources burst again. Notable examples of removed sources include SMC~X-2 (a Be X-ray pulsar) and T~CrB (a famous recurrent nova). In addition, some bright extended sources can cause false alarms due to their diffuse structures. Based on the in-flight performance, we have added two new extended sources, RCW~103 and RCW~86, into the special bright source list with specific veto radii (see Table 2).

This flexible catalog management strategy is not limited to XRBs and nova. In the future, similar operations may be applied to other classes of transient objects, such as highly active stellar flare candidates.  This ensures that the catalog continuously serves its primary function—filtering out steady and predictable background noise—without blinding the observatory to the reactivation of historically dormant, yet scientifically valuable, high-energy emitters.

\subsection{Applicability in the Future Missions}
\label{sect:applicability}

While the onboard reference catalog described in this work is tailor-made for the real-time triggering system of the EP-WXT, the underlying philosophy and development strategies possess broader value for future space missions. The methodology developed in this study offers a valuable reference with practical examples for future wide-field X-ray missions, particularly for upcoming telescopes utilizing lobster-eye optics, such as CATCH\citep{Schanne2026}, THESEUS\citep{Amati2021}, HiZ-GUNDAM \citep{Yonetoku2024, Hatsune2026}, and XTRA\footnote{\url{https://spie.org/astronomical-telescopes-instrumentation/presentation/XTRA-The-X-ray-TRansient-Array/14146-44}}. Although future missions must construct individual target lists based on specific scientific goals and unique instrument responses, valuable insights can be derived from these practical implementation steps. These include the empirical count-rate conversion formulas established across different energy bands, the two-layer database architecture that effectively isolates standard sources from exceptional bright or extended ones, and the approach of utilizing simulations to determine varying, brightness-dependent veto radii. These specific processing steps can be easily adopted by future X-ray observatories to construct autonomous triggering baselines. Furthermore, the official known X-ray source catalog generated and released by the EP mission in the future can directly serve as a robust baseline database to seed the transient search engines of these future wide-field satellites.
%========================
\section{Conclusion}
\label{sect:conclusion}

In this work, we detail the comprehensive methodology and technical implementation for constructing the onboard reference catalog of known X-ray sources for the EP-WXT. The successful development of this catalog serves as a fundamental cornerstone for the real-time transient search and autonomous triggering mechanisms of the EP mission. Through rigorous cross-matching and scientific filtering of historical data from multiple missions, we have established a robust database of 9,000 known sources, providing a reliable baseline of the X-ray sky for the EP ODPTS. 

The catalog exhibits excellent operational performance in orbit. In-flight data show that the system successfully masked 6,121 frequent triggers caused by known flaring stars from January 9, 2024, to April 30, 2026. This operational success effectively prevents unnecessary alerts and follow-up observations. Furthermore, the catalog is designed to be fully modifiable rather than hard-coded; it can be dynamically updated and refined based on subsequent in-orbit operational conditions in the future.

%========================

\begin{acknowledgements}
This work is based on the data obtained with the Einstein Probe (EP), also known as `Tianguan'. EP is led by the Chinese Academy of Sciences, in collaboration with the European Space Agency, the Max Planck Institute for Extraterrestrial Physics (Germany), and the Centre National d'Études Spatiales (France).
This work is supported by National Key R\&D Program of China No. 2025YFF0511100.
This work is supported by the National Natural Science Foundation of China (Grant Nos. 12333004, 12433005), and the Strategic Priority Research Program of the Chinese Academy of Sciences (Grant No.XDB0550200).
\end{acknowledgements}

\bibliographystyle{raa} % 调用 RAA 官方提供的 raa.bst
\bibliography{bibtex} % 您的 bib 文件名

\label{lastpage}

\end{document}